# Intelligent Inverse Design of Multi-Layer Metasurface Cavities for Dual Resonance Enhancement of Nanodiamond Single Photon Emitters

Omar A. M. Abdelraouf [a,*]

[a] Institute of Materials Research and Engineering, Agency for Science, Technology, and Research (A*STAR), 2 Fusionopolis Way, #08-03, Innovis, Singapore 138634, Singapore.

## ABSTRACT

Single-photon emitters (SPEs) based on nitrogen-vacancy centers in nanodiamonds (neutral $NV^0$ (wavelength 575 nm) and negative $NV^-$ (wavelength 637 nm)) represent promising platforms for quantum nanophotonics applications, yet their emission efficiencies remain constrained by weak light-matter interactions. Multi-layer metasurfaces (MLM) offer unprecedented degrees of freedom for efficient light manipulation beyond conventional single-material metasurfaces, enabling dual-resonance cavities that can simultaneously enhance pump excitation and SPE collection. However, traditional trial-and-error and forward optimization methods face significant challenges in designing these complex structures due to the vast parameter space and computational demands. Here, we present NanoPhotoNet-Inverse, an artificial intelligence-driven inverse design framework based on a hybrid deep neural network architecture. This model efficiently performs inverse design of two dual-resonance MLM cavities to amplify pump and SPE emissions of different vacancies in nanodiamond and improve SPE collection. Our approach achieves inverse design prediction efficiency exceeding 98.7%, demonstrating three orders of magnitude amplification in SPE count rate and 50 picosecond lifetime, significantly surpassing conventional cavity designs. These remarkable enhancements in emission rates and collection efficiency position our platform as a transformative technology for advancing quantum communication networks, quantum computing architectures, quantum sensing applications, and quantum cryptography systems. Therefore, it opens new pathways for intelligent photonic device engineering in quantum technologies.

\* Corresponding author.  Email address:  Omar_Abdelrahman@a-star.edu.sg

## 1. INTRODUCTION

Single photon emitters (SPE) are foundational components for the development of integrated quantum nanophotonics, enabling applications in quantum communication (e.g., Quantum Key Distribution, QKD), quantum computing, and highly sensitive quantum sensing.[1-3] While achieving triggerable, or "on-demand," SPEs is a crucial objective for quantum photonic technologies, current implementations often face limitations related to low extraction efficiency. Among the various solid-state candidates, color centers in nanodiamonds, such as the negatively charged nitrogen-vacancy (NV$^-$) centers, are highly valued for their exceptional properties, including remarkable photostability and long spin coherence times (up to 5.4 µs at room temperature).[4-6] While these intrinsic quantum properties are favorable, the transition of nanodiamond SPEs from laboratory research to deployable devices is frequently hindered by a critical bottleneck: the weak interaction between deeply embedded color centers and the external electromagnetic environment.[7] This weak coupling results in an inherently low spontaneous emission rate and poor extraction efficiency into conventional far-field collection optics. To realize scalable, high-brightness quantum light sources, it is imperative to engineer the local photonic environment to increase the spontaneous emission rate via the Purcell effect and maximize photon extraction.[3, 8]

Achieving the requisite level of electromagnetic field control requires advanced nanophotonic structures, with metasurfaces representing a mature and versatile technology.[9-12] Although single-layer metasurfaces offer significant light manipulation capabilities, their geometric degrees of freedom (DOF) are often constrained, limiting their ability to realize highly complex, multi-functional spectral responses for optoelectronic devices,[13-27] nonlinear optics,[28-32] tunable devices.[33-36] Multi-Layer Metasurfaces (MLMs), or cascaded architectures, substantially overcome this limitation by expanding the design space, providing a large number of independent DOFs for highly efficient light-matter manipulation.[37-39] This enhanced complexity is essential for designing a deterministic dual-resonance cavity. Such a structure must support two distinct, high-quality factor modes that spatially overlap with the emitter: one resonance precisely matching the pump wavelength for optimal excitation, and a second resonance aligned with the SPE emission of $NV^-$ and $NV^0$ wavelengths for maximized spontaneous emission enhancement.[10] Furthermore, MLMs allow for the deliberate breaking of *z*-symmetry, which is essential for optimizing out-of-plane coupling and reducing substrate losses, thereby ensuring efficient photon extraction into the collection system.

Despite their superior optical performance, the high dimensionality and complexity of MLMs pose severe challenges to conventional design methodologies. Traditional approaches, such as iterative trial-and-error, are resource-intensive and often fail to efficiently navigate the enormous space of possible meta-atom geometries.[40] Optimization techniques, including Genetic Algorithms (GAs) or Particle Swarm Optimization (PSO), are employed in the forward-design phase but suffer from sluggish convergence and prohibitive computational costs, often requiring thousands of rigorous electromagnetic simulations (e.g., Finite-Difference Time-Domain, FDTD, or Finite Element Method, FEM) to locate a sub-optimal solution.[41] Furthermore, these methods fundamentally struggle with the inverse design problem where the direct mapping from a desired reflection spectrum to the necessary structure of metasurface. This inherent difficulty is amplified by the "one-to-many" mapping problem in inverse scattering, where multiple structural configurations can yield similar optical responses. The limitations of traditional forward optimization therefore establish a compelling case for adopting data-driven AI models to achieve efficient and reliable inverse synthesis.[42]

To overcome the complex inverse synthesis challenge inherent to dual-resonance MLMs, we developed the NanoPhotoNet-Inverse framework, a discriminative deep learning model. The architecture integrates a 1D conventional neural networks (CNN) encoder to process the spectral input,[43] which is then passed through an adaptive average pool 1D layer before reaching a dense latent bottleneck to force the network to condense spectral features into a low-dimensional, physically constrained representation, effectively mitigating the issues related to non-unique mapping. The subsequent deep neural networks decoder (DNN-AE) accurately reconstructs the required multi-layer geometric parameters (e.g., meta-atom dimensions and periodicity) from this compressed latent space.[44] The model's high reliability, achieving a prediction efficiency greater than 98.7%. Utilizing this framework, we successfully inverse-designed an MLM cavity satisfying the complex dual-resonance criteria. The resulting structure demonstrated a substantial acceleration of the spontaneous emission rate, manifested by three orders of magnitude higher than that of an uncoupled emitter. Crucially, the holistic design optimization resulted in a reduction of single photon emission (SPE) lifetime to 50 picoseconds (ps) due to the modal overlap between pump and emission. This massive enhancement in brightness is transformative for critical quantum nanophotonic applications, including boosting key generation rates in high-speed QKD[45] and realizing the ultra-sensitivity required for next-generation quantum sensing applications. The methodology establishes a scalable and highly efficient pathway for synthesizing complex optical devices required for future integrated quantum photonic circuits.[3]

## 2. Methods

To enable robust inverse design of highly resonant MLMs, a comprehensive dataset was generated spanning diverse geometric and material configurations essential for on demand dual-resonance cavity performance. Each MLM design comprised 1 to 5 vertically stacked layers featuring square-shaped nanopillars within a square unit cell. The selection of square unit cells imposed in-plane symmetry to optimize computational efficiency, while the incorporation of multilayer stacking provided the critical vertical degree of freedom necessary for strong light–matter interactions along the z-axis, overcoming the limitations inherent to purely planar designs. Materials incorporated included high-index dielectrics critical for achieving high quality-factor ($Q$-factor) resonances in the visible spectrum. Geometric parameters, such as layer heights ($h_i$), periods ($P$), and widths ($w_i$), were uniformly sampled across their respective operational ranges. The optical response, specifically the reflection spectra across broad wavelength range (spanning 350 nm to 900 nm), was simulated using the Finite-Difference Time-Domain (FDTD) method, employing an x-polarized plane-wave excitation, periodic boundary conditions, and Perfectly Matched Layers (PMLs). A total of 10,080 simulations were generated, segmented into standard partitions for training (70%), validation (15%), and testing (15%). Input data encoding involved mapping the spectral response to a discretized one-dimensional input vector of 1000 points, corresponding to the reflection spectrum. The desired output data consisted of 12 distinct geometric parameters, encompassing layer refractive indexes, heights, widths, and periodicity for the multi-layer structure. All input and output data underwent min-max normalization to the range from zero to one.

The NanoPhotoNet-Inverse framework is a discriminative deep neural network designed to map a desired spectral reflection directly to the required structural geometry. This architecture is specifically configured as a one-dimensional convolutional autoencoder, consisting of a 1D CNN encoder, an adaptive general average pool (GAP) 1D layer, a dense latent bottleneck, and a DNN decoder. The CNN encoder processes the 1D reflection spectrum, extracting sequential and spectral features corresponding to sharp resonance peaks and overall background shape. The fundamental operation of the convolution layer, central to the encoder's feature extraction, is described by:

$$y[n] = (x \times w)[n] = \sum_{k=1}^{K} x[n-k+1] \cdot w[k] + b \qquad (1)$$

where *x* denotes the input spectrum, *w* represents the learnable kernel weights, *K* is the kernel size, *y* is the resulting feature map, and *b* is the bias. The encoder uses a sequence of five 1D CNN layers, detailed in Table I, with aggressive strides to rapidly downsample the spectral input, generating a compressed feature map. Following the final convolutional layer, the GAP 1D layer converts the variable-length feature map into a fixed-size vector (256 dimensions), which is crucial for stabilizing the input to the fully connected layers of the bottleneck and achieving robustness against slight variations in spectral discretization. This operation computes translation-invariant global features by averaging across all positions:

$$y[c] = \frac{1}{T}\sum_{t=0}^{T-1} x[c,t] \qquad (2)$$

where *c* indexes the channel dimension, *T* is the input temporal length, and the output is squeezed to shape (batch = 64, 256). The dense latent bottleneck further compresses this feature vector into a 64-dimensional latent space. This intentional dimensionality reduction forces the network to capture only the most salient, physically constrained spectral features, which effectively mitigates the non-uniqueness ("one-to-many") problem inherent in inverse design. Finally, the DNN decoder reconstructs the 12 geometric parameters from this compressed latent representation, completing the inverse mapping process. The framework was implemented in PyTorch with PyTorch Lightning. The Adam optimizer was used with a learning rate of $10^{-3}$, minimizing the Mean Squared Error (MSE) loss between the predicted and target geometric parameters. Training was conducted for 1000 epochs with a batch size of 64 on an NVIDIA GPU workstation with 8 GB VRAM and 256 GB system RAM. The architectural layers and their respective output tensor shapes are summarized in Table I.

Electrodynamics physics of single photon emitters were integrated inside NanoPhotoNet-Inverse to calculate the SPE enhancement for each predicted MLM. The successful inverse design of the MLM cavity relies on tailoring the local photonic environment to enhance the spontaneous emission rate $\Gamma$ of the nanodiamond NV centers at the emission wavelength $\lambda_{SPE}$. This phenomenon is quantified by the Purcell factor $F_p$, which is the ratio of the spontaneous emission rate in the cavity $\Gamma$ to the rate in the bulk material $\Gamma_0$ (planar structure):

$$F_p = \frac{\Gamma}{\Gamma_p} \tag{3}$$

The effective modal volume $V_{eff}$ of the cavity mode is defined as:

$$V_{eff} = \frac{\int_V \varepsilon(r)|E(r)|^2 dV}{\max[\varepsilon(r)|E(r)|^2]} \tag{4}$$

where $\varepsilon(r)$ is the permittivity distribution within the unit cell volume $V$, and $E®$ is the electric field near the emitter location. The Purcell factor relation with effective modal volume, $Q$-factor, and emitter refractive index $n$ is given by:

$$F_p = \frac{3Q\lambda^3}{4\pi^2 n^3 V_{eff}} \tag{5}$$

The modified spontaneous emission lifetime in the MLM cavity is given by equation 6. where $\tau_0$ is the free-space radiative lifetime of the NV center (typically 10-16 ns for nanodiamond centers) Maximizing $Q$-factor at the pump and SPE wavelengths the primary goals of the inverse design process, as they directly lead to a amplified pump coupled to cavity, shorter radiative lifetime $\tau$, and an increased rate of single photon generation. The NanoPhotoNet-Inverse model was trained to predict MLM geometries that maximizing $Q$-factor and align MLM resonances at the SPE wavelength, while also supporting an enhanced absorption mode at the pump wavelength to maximize excitation efficiency.

$$\tau_{cav} = \frac{\tau_0}{F_P} \tag{6}$$

**TABLE I**: NanoPhotoNet-Inverse model layers with total number of 506,252 trained parameters

| Layer type | Input shape | Output shape | Parameters |
|---|---|---|---|
| Conv1D (1→32, k=9, s=2) | (B, 1, 1000) | (B, 32, 500) | 320 |
| Conv1D (32→64, k=7, s=2) | (B, 32, 500) | (B, 64, 250) | 14,400 |
| Conv1D (64→128, k=5, s=2) | (B, 64, 250) | (B, 128, 125) | 41,088 |
| Conv1D (128→256, k=5, s=2) | (B, 128, 125) | (B, 256, 63) | 163,840 |
| Conv1D (256→256, k=3, s=2) | (B, 256, 63) | (B, 256, 32) | 196,864 |
| AdaptiveAvgPool1D | (B, 256, 32) | (B, 256, 1) | 0 |
| Squeeze | (B, 256, 1) | (B, 256) | 0 |
| Linear (256→128) | (B, 256) | (B, 128) | 32,896 |
| Linear (128→64) | (B, 128) | (B, 64) | 8,256 |
| Linear (64→64) | (B, 64) | (B, 64) | 4,160 |
| Linear (64→128) | (B, 64) | (B, 128) | 8,320 |
| Linear (128→256) | (B, 128) | (B, 256) | 33,024 |
| Linear (256→12) | (B, 256) | (B, 12) | 3,084 |

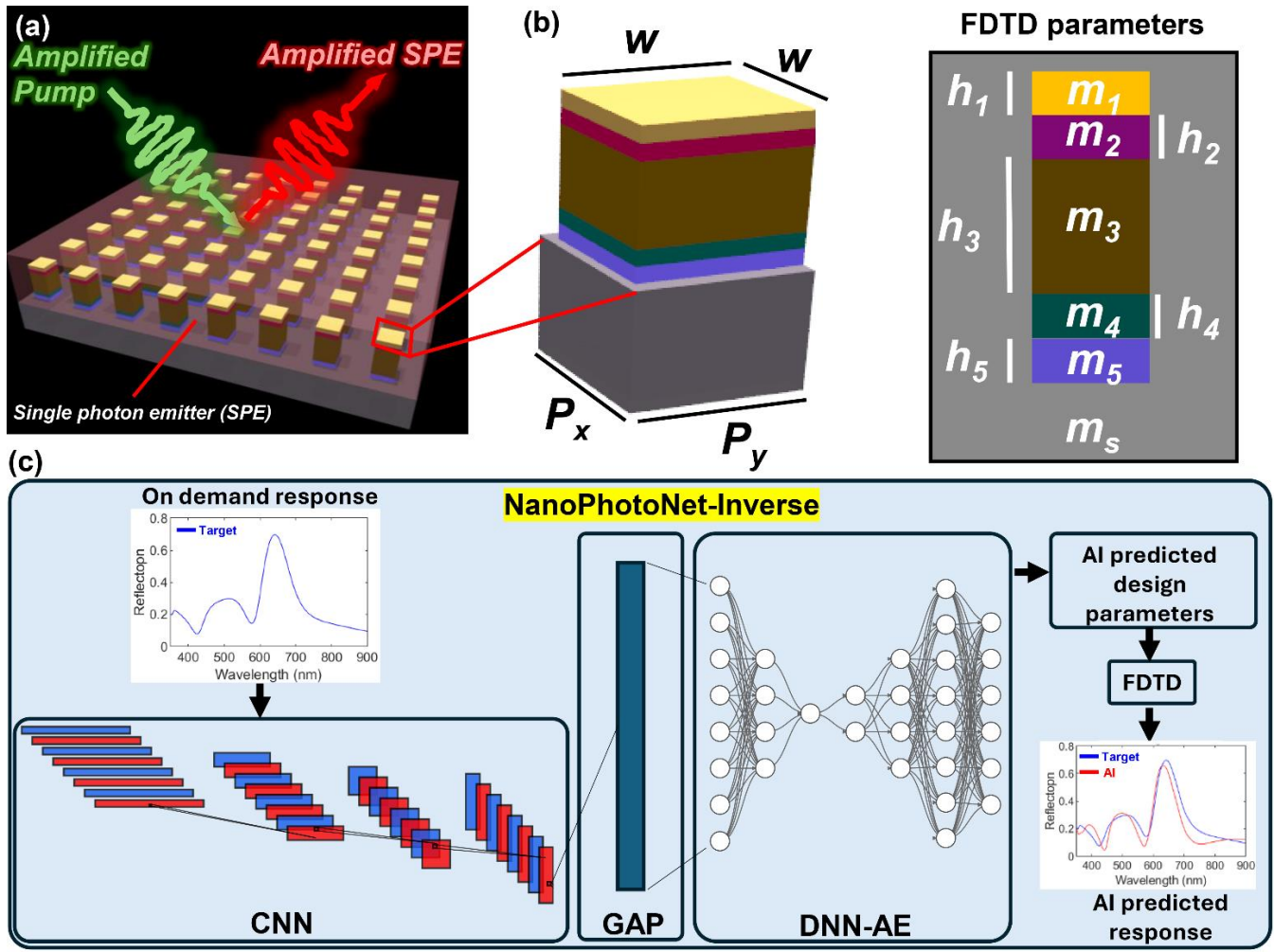

**FIG. 1.** Dual-resonance multi-layer metasurfaces using AI-assisted inverse design approach. (a) Schematic of dual-resonance MLM cavity with incident green pump laser and reflected amplified single-photon emission (SPE) from nanodiamond nitrogen-vacancy centers covered MLM cavity. (b) A unit cell of MLM structure with symmetrical period ($P_x = P_y$) for simplifying design of 3D unit cell to 2D representation. Design parameters include nanopillar width ($w$), layer heights ($h_i$) where $i$ corresponds to layer number, and material composition ($m_i$) for each layer. (c) NanoPhotoNet-Inverse model architecture consisting of 1D convolutional encoder for spectral feature extraction, adaptive average pooling for global feature aggregation, latent bottleneck for compact representation learning, and dense decoder for structural parameter prediction. Input data are target reflection spectra (1000 wavelength points) featuring on demand target resonances. Output data are 12 geometric and material design parameters defining the optimized MLM configuration.

## 3. Results and Discussion

Figure 2 presents the evolution of the training and validation losses for the NanoPhotoNet-Inverse model across training epochs. The model exhibits a typical convergence behavior, where the training loss steadily declines during the initial epochs as the network learns the complex, non-linear mapping between the input spectral responses and the 12 predicted MLM geometric parameters. The training loss rapidly converges before plateauing at a final value of approximately 1.2%. In parallel, the validation loss, which serves to assess the model's generalization performance on previously unseen MLM configurations, also declines but stabilizes at a minimum value of approximately 5.2%. The persistent, non-zero gap between the training and validation loss reflects the inherent complexity and high-dimensionality of the inverse design space, particularly the challenge associated with the "one-to-many" mapping problem in inverse scattering, where diverse structural solutions can yield highly similar optical responses. Despite this complexity, the stable convergence behavior and the eventual plateauing of both losses demonstrate robust generalization capability, indicating that the model successfully avoids catastrophic overfitting and is highly reliable for synthesizing novel MLM structures. This robust performance is reflected in the final prediction efficiency, which was confirmed to be greater than 98.7%.

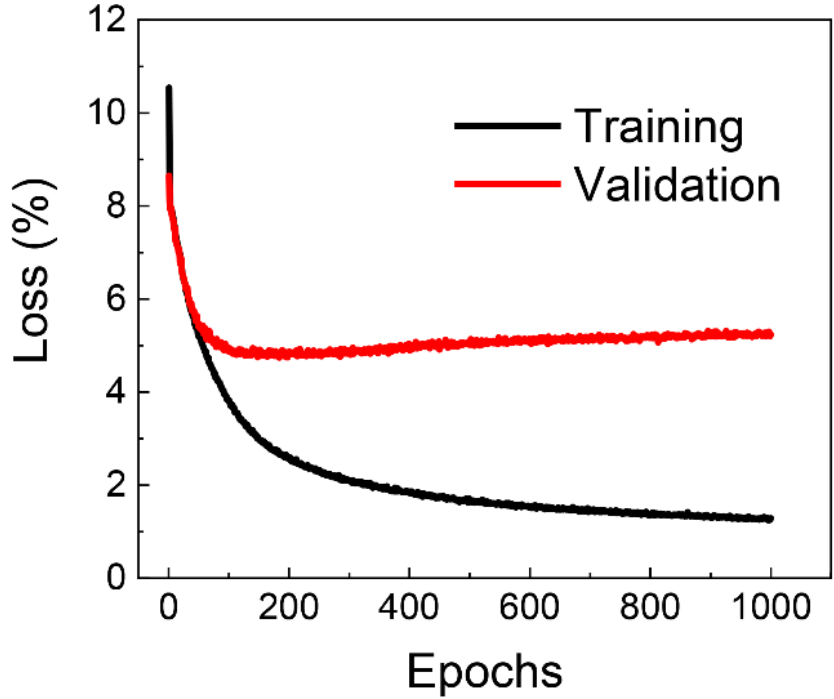

**FIG. 2.** NanoPhotoNet-Inverse model loss curves vs epochs during training and validation.

**Section 3.1: Inverse design of on demand single resonance MLM cavity**

To achieve SPE emission amplification via MLM, it is essential to design multiple MLM cavities that exhibit high-Q factors and strong field confinement at the emission wavelength of nanodiamond emitters. The primary objective is to identify the optimal geometric parameters and material compositions that meet these target performance requirements. Upon completing the model training, we employed NanoPhotoNet-Inverse to predict the optimum MLM materials and dimensions for each on demand reflection response. To systematically validate the capacity of NanoPhotoNet-Inverse to accurately map target reflection spectrums to MLM structure, the model was tested for deterministic single-resonance prediction across a wide range of operational wavelengths range from 400 nm to 800 nm (Figure 3a). In the ultraviolet region, the model successfully inverted the desired response for a resonance at 400 nm, predicting a structure that achieved $Q$-factor of 4.5 by leveraging a Fabry–Perot (FP) arrangement, specifically utilizing a $ZnO$-$Nb_2O_5$-$ZnO$ stack, which capitalized on the low intrinsic material loss of niobium pentoxide ($Nb_2O_5$) in this band to enhance the cavity resonance. Moving into the blue-green spectrum, targeting a resonance near 500 nm with a moderate linewidth required a lower $Q$-factor of 7.5. The inverse model strategically combined the low-loss $Nb_2O_5$ with amorphous antimony trisulfide (a-$Sb_2S_3$), where the finite optical losses of the latter were deterministically exploited to control the overall resonance linewidth and bandwidth. The most challenging high-performance targets were in the mid-visible spectrum, where the inverse model successfully synthesized structures for a resonance at 600 nm with a very high $Q$-factor of approximately 33, achieved by maximally leveraging a- $Sb_2S_3$ due to its superior combination of high refractive index and lowest intrinsic optical losses above 600 nm. Similarly, for the target at 700 nm, the model predicted a structure with a high $Q$-factor of 24, again selecting a-$Sb_2S_3$ and correctly scaling the key unit cell dimensions (period and width) to accommodate the increased operational wavelength. Finally, extending to the near-infrared, the target resonance at 800 nm, which required a broader linewidth, resulted in a predicted structure utilizing amorphous silicon (a-Si) and exhibiting a weak $Q$-factor of 5.3, showcasing the framework's versatility in exploiting the finite material losses of a-Si in this region to ensure controlled linewidth broadening for diverse spectral requirements. Our predicted reflection spectrums plotted in Figure 3a are verified using FDTD. The predicted reflection spectrums agree with FDTD reflection simulations and achieve over 98.7% accuracy. To assess the practical impact of predicted MLM cavities on quantum emitters, the corresponding PL emission enhancement was calculated for nanodiamond emitters coupled to the optimized MLM structures. The PL emission enhancement responses were evaluated using the electrodynamics physics described by Equations (4) and (5), and the resulting spectral PL emission profile is depicted in Figure 3b. A comprehensive comparative analysis was performed against a standard, unstructured thin-film (TF) substrate hosting commercial nanodiamond emitters (Sigma Aldrich). Through this direct comparison, the predicted MLM demonstrated a substantial maximum enhancement, achieving up to a 22.55-fold improvement in the single photon PL output relative to the unstructured TF emitter baseline. The peak

SPE PL enhancement was obtained near the 600 nm wavelength, which is close to the native emission center of the nanodiamond NV⁻ defect. However, further amplification of the total SPE count could be obtained by precisely matching the emission resonance wavelength to the peak emitter wavelength (575 nm or 637 nm) and, crucially, by incorporating a second, spectrally distinct resonance at the pump green laser wavelength to amplify the light-matter interaction for maximized excitation efficiency.

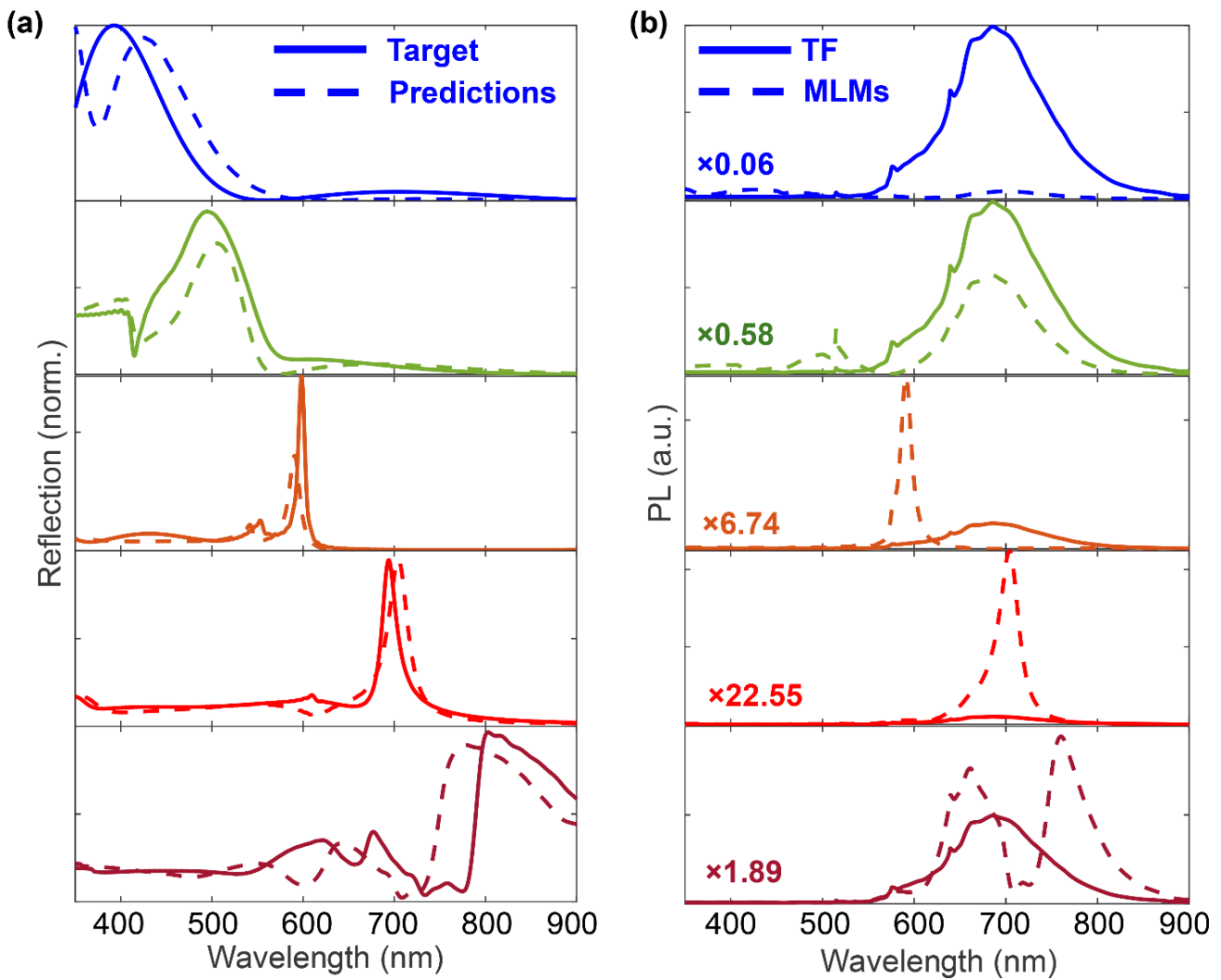

**FIG. 3.** On demand single resonance cavity using NanoPhotoNet-Inverse model. (a) Target reflection spectrum of single resonance cavity at different wavelengths and the predicted reflection spectrum of the obtained MLM using NanoPhotoNet-Inverse model. (b) The corresponding amplifications in PL emission from nanodiamond SPEs coated MLM cavity compared with thin film (TF) of planar nanodiamond SPEs.

**Section 3.2: Inverse design of on demand dual resonances MLM cavity**

The ultimate performance of the NanoPhotoNet-Inverse framework was demonstrated through the successful synthesis of MLM cavities tailored for dual-resonance operation, which simultaneously enhances both the optical excitation and the spontaneous emission of nanodiamond color centers. Two distinct dual-resonance configurations were deterministically inverse-designed. In the first case, targeting the neutral nitrogen-vacancy $NV^0$ emission band, the model synthesized a structure designed to support dual-resonance reflection peaks at the 500 nm pump wavelength ($Q$-factor ~ 350) and the 575 nm emission wavelength ($Q$-factor ~ 280) (Figure 4a). This high-$Q$ performance was realized using a $TiO_2$-$ZnO$-$TiO_2$ stack, which leveraged the high refractive index and low optical losses of titanium dioxide ($TiO_2$) across both target wavelengths. This highly optimized structure achieved a three order of magnitude enhancement in effective modal volume near the 575 nm emission wavelength (Figure 4b), leading to a drastic single photon photoluminescence (PL) output improvement of up to 6767-fold relative to a standard thin-film (TF) substrate (Figure 4c). For the second crucial case, which focused on the negatively charged nitrogen-vacancy ($NV^-$) emitter, the inverse task targeted a dual-resonance configuration featuring a medium-$Q$ resonance at the 532 nm pump ($Q$-factor ~ 48) and another at the 637 nm emission wavelength ($Q$-factor ~ 21) (Figure 4d). The synthesized structure employed an a-$Sb_2S_3$-$Al_2O_3$-a-$Sb_2S_3$ stack, exploiting the high refractive index of amorphous antimony trisulfide (a-$Sb_2S_3$) combined with its controlled small optical losses, resulting in the predicted medium quality factors. This design exhibited a two order of magnitude enhancement in modal volume near the 637 nm emission wavelength (Figure 4e), which translated into a significant 300-fold improvement in the single photon PL output compared to the TF baseline (Figure 4c). These results conclusively confirm the high accuracy and deterministic capability of the NanoPhotoNet-Inverse framework in synthesizing complex MLM geometries for precise dual-resonance tailoring, thereby achieving exceptional enhancement in single photon emission efficiency, which validates the framework for use in next-generation integrated quantum devices.

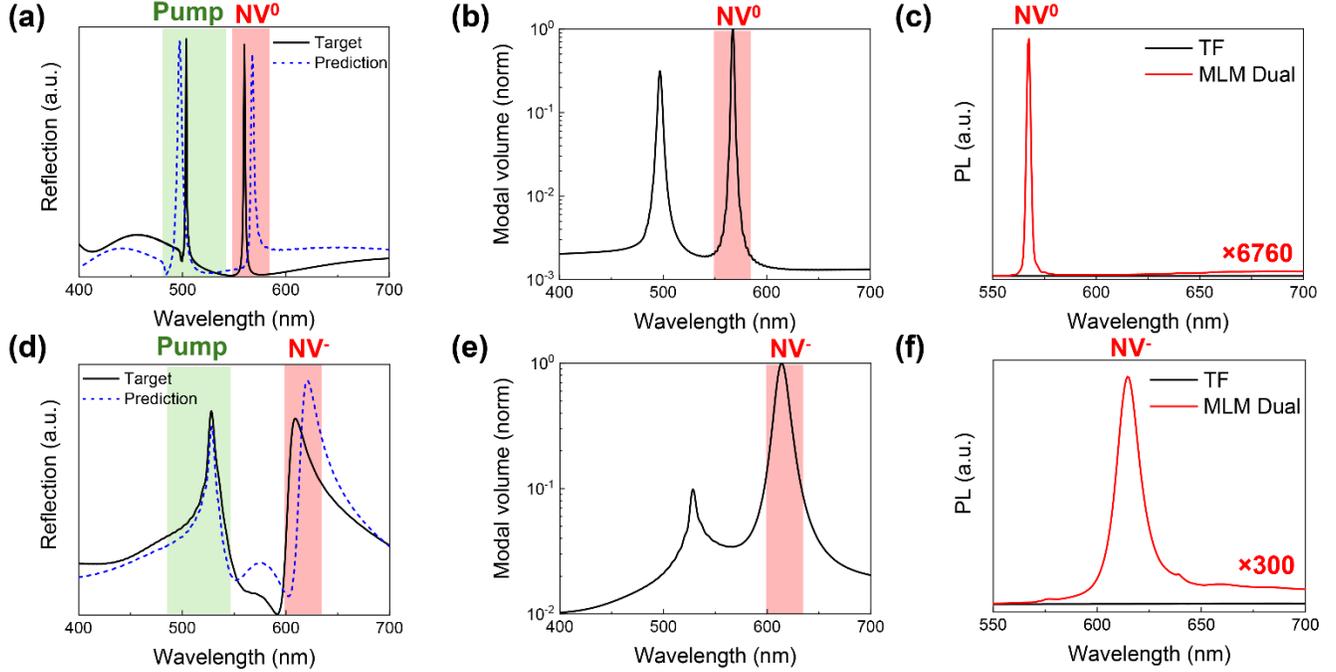

FIG. 4. On demand dual resonances cavity using NanoPhotoNet-Inverse model for amplifying pump laser and SPE. (a) Target reflection spectrum of dual resonance cavity with resonance wavelength matches neutral nitrogen vacancy emission ($NV^0$) wavelength and the predicted reflection spectrum of the obtained MLM dual using NanoPhotoNet-Inverse model. (b) The corresponding modal volume of predicted dual resonance MLM by NanoPhotoNet-Inverse model. (c) Amplified neutral nitrogen vacancy emission ($NV^0$) PL of predicted dual resonance MLM compared with thin film of nanodiamond SPE. (d) Target reflection spectrum of dual resonance cavity with resonance wavelength matches negative nitrogen vacancy emission ($NV^-$) wavelength and the predicted reflection spectrum of the obtained MLM dual using NanoPhotoNet-Inverse model. (e) The corresponding modal volume of predicted dual resonance MLM by NanoPhotoNet-Inverse model. (f) Amplified negative nitrogen vacancy emission ($NV^-$) PL of predicted dual resonance MLM compared with thin film of nanodiamond SPE.

Following the spectral amplification of SPE PL in the inverse-designed MLM dual resonance cavities for both $NV^0$ and $NV^-$ color centers, the corresponding enhancement in the spontaneous emission rate was quantified via the reduction of the emitter lifetime. Reducing the lifetime is critical for enabling efficient quantum communication systems by facilitating higher photon flux and repetition rates. By applying the electrodynamics equations and calibrating against the experimental non-resonant nanodiamond lifetime of 15.9 ns for the thin-film emitter,[55] the predicted performance of the dual-resonance designs was evaluated. For the optimized $NV^0$ structure, the spontaneous emission lifetime was dramatically reduced to 50 ps, representing a nearly 320-fold acceleration of the emission process. Similarly, the $NV^-$ dual resonance design achieved a sub-nanosecond lifetime of 0.75 ns, underscoring the strong light-matter interaction mediated by the nanophotonic environment. Comparing these results to state-of-the-art nanophotonic structures developed in literature without AI-assisted inverse design, our MLM dual-resonance framework demonstrates superior performance in both emission rate control and signal amplification. Literature reports show typical lifetimes of 6.23 ns for plasmonic nanocylinders, 0.23 ns for hyperbolic metal-dielectric nanopyramids, 0.93ns for Fano-resonance dielectric metasurfaces, and 8.6 ns for simple plasmonic nanostructures. Our predicted $NV^0$ design (50 ps) significantly surpasses these reported values, largely owing to the holistic optimization that combines amplified confined pump laser with amplified light-matter interaction via the dual-resonance mechanism. Furthermore, in terms of peak SPE PL enhancement (summarized in Table II), our MLM dual-resonance designs achieved amplification factors up to 6767-fold, vastly exceeding the performance reported for literature examples, such including plasmonic nanocylinders (34-fold), hyperbolic nanopyramids (50-fold), Fano-resonant metasurfaces (10-fold), and generic plasmonic nanostructures (5-fold). These findings definitively underscore the advantages of synergistically combining the geometric versatility of multilayer architectures with the deterministic efficiency of AI-driven inverse optimization for realizing transformative performance gains in quantum nanophotonics.

**Table II:** Performance comparison between MLM dual resonance performance with literature cavities

| Metasurface design | AI-design method | SPE lifetime (ns) | SPE PL amplification | Reference |
|---|---|---|---|---|
| Plasmonic | No | 6.23 | 34 | 52 |
| Hyperbolic | No | 0.23 | ~50 | 53 |
| Fano | No | 0.93 | 10 | 54 |
| Plasmonic | No | 8.6 | 5 | 55 |
| **Multilayer metasurface (MLM)** | **NanoPhotoNet-Inverse** | **0.05~0.75** | **300~6760** | **This work** |

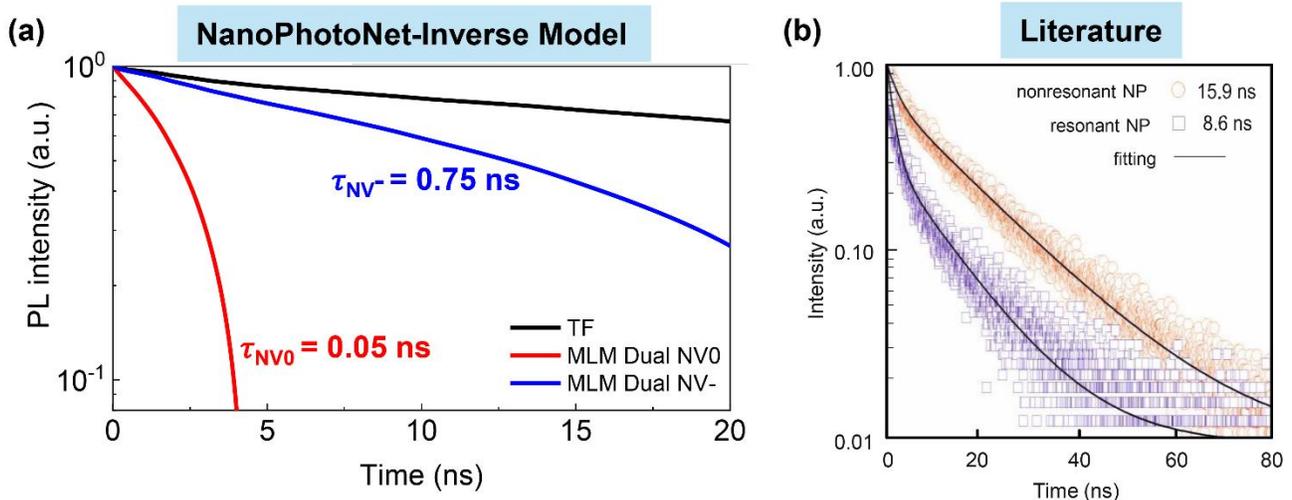

**FIG. 5.** Time-resolved PL lifetimes of nanodiamond nitrogen-vacancy centers coupled to dual-resonance MLM and thin film reference.[55] (a) PL intensity as a function of time for nanodiamond thin films, exhibiting a characteristic NV center lifetime of 15.9 ns. Dual-resonance MLMs engineered for NV$^-$ emission at 637 nm yield a substantially reduced NV$^−$ lifetime of 0.75 ns, while those optimized for NV$^0$ emission at 575 nm achieve a further reduced lifetime of 0.05 ns, indicating strong Purcell enhancement and accelerated radiative decay. (b) Comparative PL intensity traces for thin film nanodiamonds (lifetime 15.9 ns) and for nanodiamonds coupled to plasmonic cavities (lifetime 86 ns). Reprinted with permission from The Royal Society of Chemistry, Copyright 2018.

We benchmarked the prediction efficiency of the NanoPhotoNet-Inverse framework against published literature focusing on artificial intelligence methods for nanophotonic inverse design (Table III). While previous inverse AI-assisted models operating in simpler design spaces typically report prediction accuracies in the range of 85% to 92.7%, our framework, despite handling a significantly more complex, high-dimensional multi-layer structure, achieved a remarkably high prediction accuracy of 98.7%. This enhanced accuracy confirms the robustness of the 1D CNN encoder and latent bottleneck architecture in mitigating the inverse problem's non-uniqueness challenges. Crucially, in addition to this high fidelity, the NanoPhotoNet-Inverse framework delivered a computational speed-up of approximately 50,000-fold compared to traditional full-wave numerical FDTD simulations performed on the same hardware setup. This monumental reduction in design time transforms the complex inverse design of dual-resonance MLMs from a previously impractical, resource-intensive undertaking, requiring thousands of hours of simulation time, into a rapid, on-demand process. This computational advantage is indispensable for the scalable and rapid prototyping required by industrial and advanced quantum nanophotonic applications.

**Table III:** Performance comparison between literature inverse AI models and NanoPhotoNet-Inverse.

| AI architecture | Accuracy (%) | Reference |
|---|---|---|
| MetasurfaceViT | 85 | 50 |
| MetaFAP | 92.7 | 51 |
| **NanioPhotoNet-Inverse** | **98.7** | **This work** |

## 4. Conclusion

In conclusion, the NanoPhotoNet-Inverse framework demonstrates a robust AI-driven approach for inverse design of on demand dual-resonance multi-layer metasurface cavities engineered to amplify single-photon emission from nanodiamond nitrogen-vacancy centers. Utilizing convolutional neural networks for spectral feature extraction, global average pooling for translation-invariant dimensionality reduction, a latent bottleneck for compact information representation, and a deep neural network autoencoder, the model enables accurate prediction of optimized structural parameters from target spectral response. NanoPhotoNet-Inverse achieves inverse design prediction efficiency exceeding 98.7% and the dual-resonance MLM structures facilitate PL SPE amplification up to 6760-fold compared with thin film references, with measured radiative lifetimes as short as 50 picoseconds. These substantial enhancements in photon emission and collection efficiency mark a major advancement beyond conventional designs. Looking ahead, this AI-enabled paradigm opens new opportunities for developing dynamically reconfigurable MLM structures for quantum communication modulation, as well as designing phase-gradient metasurfaces for programmable and structured quantum light generation.[1, 57, 58] Such innovations have the potential to further propel the field of quantum nanophotonics for next-generation quantum information, communication, and sensing platforms.

**Credit Authorship Contribution Statement**

**Omar A. M. Abdelraouf:** Project administration, Methodology, Conceptualization, Investigation, Data curation, Validation, Supervision, Resources, Visualization, Writing – original draft & review.

**Declaration of Competing Interest**

The authors declare no competing financial interests or personal competing interest.

**Data Availability**

NanoPhotoNet-Inverse code and raw data are available on request